\documentclass[runningheads]{svmult}
\usepackage{makeidx}  
\usepackage{graphicx}  
\usepackage{subeqnar}  
\usepackage{multicol} 
\usepackage{cropmark} 
\usepackage{physprbb}

\begin{document}
\title*{Higgs Bosons and the Indirect Search for WIMPs}
\author{V. A. Bednyakov\inst{2} 
\and    H. V. Klapdor-Kleingrothaus\inst{1}
\and    H. Tu\inst{1}}

\institute{Max-Planck-Institut f\"{u}r Kernphysik, 
      Postfach 103980, D-69029 Heidelberg, Germany
\and  Laboratory of Nuclear Problems,
      Joint Institute for Nuclear Research,
      Moscow region, 141980 Dubna, Russia}
 
\maketitle

\begin{abstract}
        We investigated the contribution of the MSSM Higgs bosons 
        produced in the 
        neutralino annihilation in the Earth and Sun 
        to the total WIMPs detection signal.
        We found that this contribution is very important and 
        results in a lower bound for the muon flux from the Sun of  
        $10^{-7} \div 10^{-8}$ m$^{-2}$ yr$^{-1}$ 
        for neutralinos heavier than about 200 GeV.
        We noticed that due to the SUSY charged Higgs bosons 
        one can expect an energetic $\tau$ neutrino flux from the
        Sun at a level of $10^2$ m$^{-2}$ yr$^{-1}$.
\end{abstract}

\section{Introduction}
        The Weakly-Interacting Massive Particles (WIMPs) 
        as the most attractive and well-motivated cold dark-matter candidates 
        in the Galactic Halo may be gravitationally trapped 
        by astrophysical objects like the Earth or the Sun due to
        subsequent energy losses via elastic 
        scatterings off the nuclei therein. 
        The trapped WIMPs would undergo pair annihilations in the core 
        of the Earth or the Sun, producing all kinds of ordinary particles 
        like quarks, leptons, gauge and Higgs bosons.
        The neutrinos produced in the decays of these WIMP 
        annihilation products provide detectable signals, because among 
        all particles from WIMPs annihilation  only neutrinos can pass 
        through the astrophysical bodies and reach the detectors. 
        The flux of the neutrinos of type $i$ (e.g. $i=\nu_{\mu}$, 
        $\bar{\nu}_{\mu}$, or $\nu_{\tau}$, $\bar{\nu}_{\tau}$) 
        from WIMP annihilation in the Earth or Sun is
\cite{Jun96}
\begin{equation}
   \left(\frac{d \phi}{d E} \right)_i = \frac{\Gamma_A}{4 \pi R^2}
   \sum_F B_F \left(\frac{d N}{d E} \right)_{F,i} 
\label{flux-spectrum}
\end{equation}
        where $\Gamma_A$ is the annihilation rate of the WIMPs 
        in the Sun or Earth, $R$ is the distance the 
        neutrinos have to travel to the detector, 
        $B_F$ is the branching ratio for annihilation into final state $F$, 
        and $(d N /d E)_{F,i}$ is the differential energy spectrum of 
        the $i$-type neutrinos at the surface of the Sun or Earth 
        expected from decays of particles produced in channel $F$ 
        in the core of the Sun or Earth.
        Recent information about indirect WIMP detection
        can be found in e.g.
\cite{Jun96,suv99,Bottino,Bergstrom},
        or in the talks given by T. Montaruli, O. Suvorova, 
        J. A. Djilkibaev, N. Fornengo, A. Bottino at  
        this conference.

        Among all WIMPs, the neutralinos are the most considered, 
        which arise from the supersymmetric extensions of the 
        Standard Model and are most likely the 
        Lightest Supersymmetric Particle (LSP).
        In SUSY scenarios with large sfermion and gaugino masses 
        even the LSP should also be heavy and thus cannot be produced at 
        current or near-future colliders.
        In this case the Higgs bosons still have chances to be light and 
        specifically the detection of the charged Higgs boson could 
        serve as first evidence for supersymmetry. 
 
        With this interest in the Higgs bosons, 
        in this work we perform an investigation within the framework of 
        the Minimal Supersymmetric extension of the Standard Model 
        on the Higgs bosons produced in the neutralino 
        annihilation in the Earth and Sun and their role in 
        the detection of the WIMPs.  
        Following our approach in
\cite{approach},  we relax any constraints from unification
        assumptions on the parameters.
        On the other hand we use the results from 
        collider searches for supersymmetric particles, 
        rare FCNC $b \rightarrow s \gamma$ decay, 
        as well as bounds on the WIMPs relic abundances
        to constrain the MSSM parameter space.
        Our free parameters are 
        $\tan \beta$, $\mu$, $M_1$, $M_2$, $M_A$, $m^2_{\tilde{Q}}$,
        $m^2_{\tilde{L}}$,  
        $m^2_{\tilde{Q_3}}$, 
        $m^2_{\tilde{L_3}}$ and $A_t$
(see \cite{approach}).

\section{Higgs Boson Contribution to Indirect WIMP Search}
        The Minimal Supersymmetric extension of the Standard Model (MSSM) 
        possesses two complex scalar doublets, after the spontaneous 
        electroweak-symmetry breaking five physical Higgs
        bosons appear in the MSSM particle spectrum.
        Assuming $CP$ invariance, the MSSM Higgs sector contains a neutral 
        $CP$-odd Higgs boson $A^0$, two neutral $CP$-even Higgs bosons $H^0$, 
        $h^0$, which are mixtures of the neutral Higgs interaction 
        eigenstates, and a pair of charged Higgs bosons $H^ {\pm}$.
        All the five MSSM Higgs bosons can be produced in neutralino pair 
        annihilation processes, either in pairs or accompanied by 
        a gauge boson.
        However, since the pair annihilation of the neutralinos captured 
        in the Sun and Earth takes place practically at rest,
        only processes allowed by $CP$ conservation in the 
        $v \rightarrow 0$ limit 
        are relevant for the indirect WIMPs detection:
\begin{equation}
   \chi \chi \rightarrow h^0 Z^0,\ h^0 A^0,\ H^0 Z^0,\ H^0 A^0,\
   W^{\pm} H^{\mp}.
\end{equation}

The Higgs bosons produced in neutralino annihilation decay immediately
before losing energy.
Because the Higgs couplings to fermion pairs are proportional to
the fermion masses (see for example 
\cite{Higgs-Hunters}),
the neutral Higgs bosons do not decay into neutrinos directly,
and the branching ratio for charged Higgs boson decay into muon
neutrino is negligible.
Also, for the Higgs bosons produced in the LSP neutralino annihilation
at rest the decay into superpartners is kinematically forbidden.
Therefore we consider the following two-body decay channels of the five 
MSSM Higgs bosons which can produce energetic muon neutrinos
\begin{eqnarray}
   h^0 &\rightarrow& \tau \bar{\tau}, 
   \hspace{0.1cm}b \bar{b}, 
   \hspace{0.1cm}c \bar{c}, 
   \hspace{0.1cm}t \bar{t}, 
   \hspace{0.1cm}W^+ W^-, 
   \hspace{0.1cm}Z^0 Z^0, 
   \hspace{0.1cm}Z^0 \gamma; 
   \nonumber \\
   H^0 &\rightarrow& \tau \bar{\tau}, 
   \hspace{0.1cm}b \bar{b}, 
   \hspace{0.1cm}c \bar{c}, 
   \hspace{0.1cm}t \bar{t}, 
   \hspace{0.1cm}W^+ W^-, 
   \hspace{0.1cm}Z^0 Z^0, 
   \hspace{0.1cm}Z^0 \gamma,
   \hspace{0.1cm}Z^0 A^0, 
   \hspace{0.1cm}h^0 h^0, 
   \hspace{0.1cm}A^0 A^0, 
   \hspace{0.1cm}W^{\pm} H^{\mp}; 
   \nonumber \\
   A^0 &\rightarrow& \tau \bar{\tau},
   \hspace{0.1cm}b \bar{b}, 
   \hspace{0.1cm}c \bar{c}, 
   \hspace{0.1cm}t \bar{t}, 
   \hspace{0.1cm}Z^0 \gamma, 
   \hspace{0.1cm}Z^0 h^0; 
   \nonumber \\
   H^+ &\rightarrow& \tau^+ \nu_{\tau},
   \hspace{0.1cm}c \bar{b}, 
   \hspace{0.1cm}c \bar{s}, 
   \hspace{0.1cm}t \bar{b}, 
   \hspace{0.1cm}h^0 W^+, 
   \hspace{0.1cm}A^0 W^+ 
        \quad + \quad {\rm charge~conjugation}.
\end{eqnarray}
        The decay branching ratios for the above channels 
        are calculated with the computer code
        HDECAY written by A. Djouadi et al. 
\cite{HDECAY}, in which all relevant higher-order QCD corrections to 
        decays into quarks and gluons, all important below-threshold 
        three-body decays, and complete radiative corrections in the 
        effective potential approach are included.

        We investigate the Higgs boson contribution 
        to the WIMPs detection signal
(\ref{flux-spectrum}) by switching off the final states 
        containing Higgs bosons in our calculation 
        (that is, we do not count the neutrinos from these channels)
        and compare the results 
        $\phi_{\rm no Higgs}$ 
        with that from all channels including Higgs final states 
        $\phi_{\rm with Higgs}$, 
        which is calculated using the formulae given in
\cite{Jun96}.

\begin{figure}[h]
\begin{center}
\includegraphics[height=11cm]{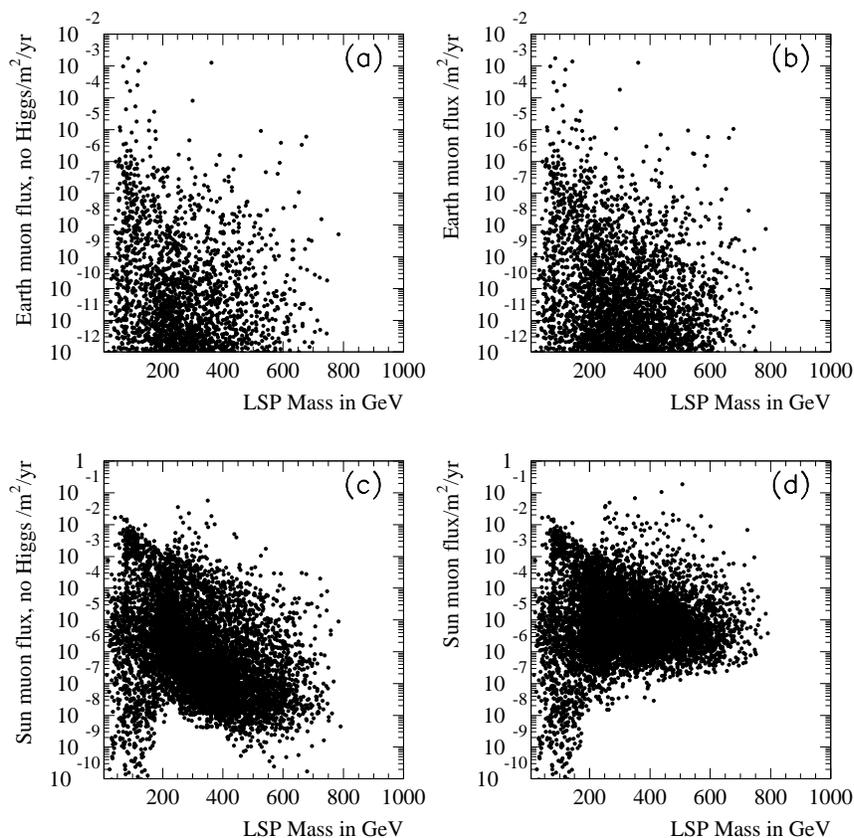} 
\end{center}
\caption[]{Upward-going muon detection rate from the Earth 
        (a) without contribution from final states containing Higgs bosons 
        $\phi^{\oplus}_{\rm no Higgs}$ and 
        (b) with all channels including Higgs final states
        $\phi^{\oplus}_{\rm with Higgs}$  
        as well as from the Sun
        (c) without $\phi^{\odot}_{\rm no Higgs}$, 
        (d) with $\phi^{\odot}_{\rm with Higgs}$ 
        Higgs final states.
}\label{higgs1}
\end{figure}

\begin{figure}[h]
\begin{center}
\includegraphics[height=11cm]{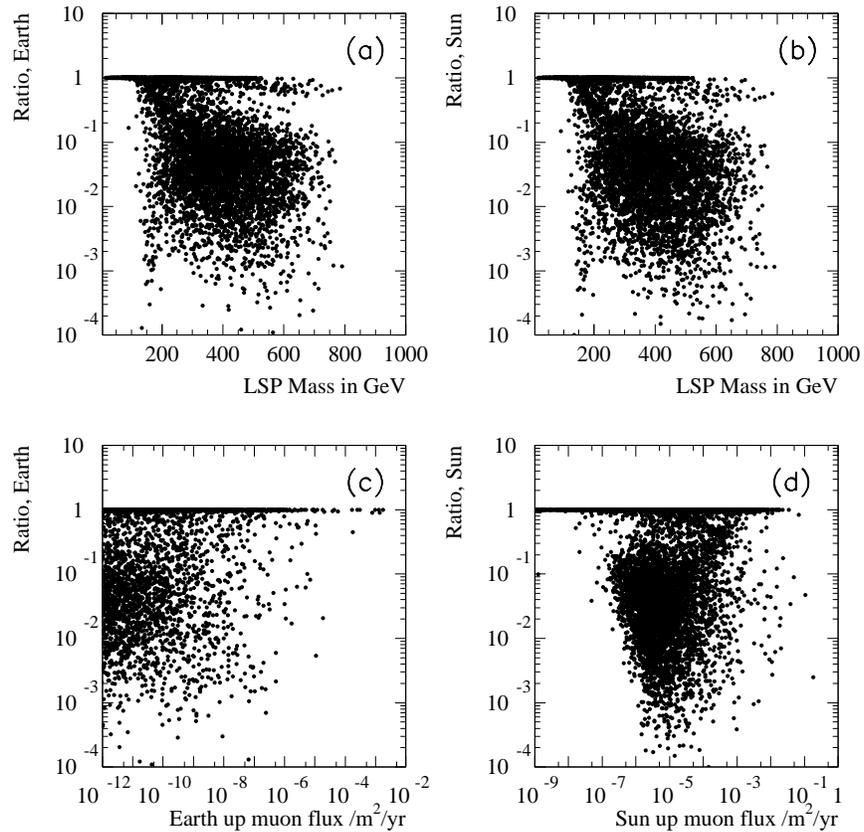} 
\end{center}
\caption[]{Ratio of the detection rate without Higgs final states to from
all channels including Higgs final states 
(a) and (b) versus $m_{\chi}$, (c) and (d) versus the detection
rate $\phi_{\rm with Higgs}$.}
\label{higgs2}
\end{figure}

        In figure 
\ref{higgs1}
        we plot our results for 
        $\phi_{\rm no Higgs}$ and $\phi_{\rm with Higgs}$ 
        versus the neutralino mass $m_{\chi}$.
Comparing figure 
\ref{higgs1}(c) with 
\ref{higgs1}(d) we see 
        (it can be seen more easily in the case of the Sun), 
        that there are significant differences between   
        them mainly in the region where the total detection rate for 
        upward-going muons induced by neutrinos from the Sun is
        $10^{-7} \div 10^{-9}$ m$^{-2}$ yr$^{-1}$
        for neutralinos heavier than about 200 GeV.   
        In order to see the Higgs final state contribution 
        quantitatively we plot in figure 
\ref{higgs2} 
        the ratio of the two detection rates
        $\phi_{\rm no Higgs} / \phi_{\rm with Higgs}$ (a), (b) 
        versus neutralino mass $m_{\chi}$ and (c), (d) versus the 
        upward-going muon detection rate 
        with contribution from all channels $\phi_{\rm all}$. 

        In figure 
\ref{higgs2}(a) and 
\ref{higgs2}(b) it can be seen that for neutralinos heavier than 200 GeV
        the Higgs final states are in general important. 
        If their contributions were not included 
        the indirect WIMP detection rate could be roughly 
        $10^1$ to $10^3$ times lower. 
        Figure 
\ref{higgs2}(d) shows that in the case of the Sun, the Higgs final state 
        contribution is important in the region 
        $10^{-7} \leq \phi^{\odot}_{\rm all} \leq 10^{-3}$ m$^{-2}$ yr$^{-1}$
        where one should expect the signals.

        The Higgs contribution is not important in the region where the 
        upward-going muon detection rate from the Sun is lower than 
        $10^{-7}$ m$^{-2}$ yr$^{-1}$, because such low detection rates 
        are expected mainly for neutralinos lighter than about 200 GeV 
(see figure \ref{higgs1}(d)).
        In this case the neutralino annihilation into final states 
        containing Higgs bosons may have chances to compete with, or even 
        dominate over the fermion and/or gauge boson pair final states 
        if kinematically allowed, 
        but since the Higgs bosons do not produce neutrinos directly, 
        the neutrinos from the Higgses are not so energetic compared 
        to those directly produced in the fermion or gauge boson decay.
        As a consequence, their contribution to the total detection rate 
        which is proportional to the product of their branching ratios 
        $B_F$ and the second moment of the neutrino spectra 
        $\left( d N /d E \right)_{F,i}$, 
        is not important for lighter neutralinos.

        The indirect dark matter search experiments have now reached 
        a sensitivity of $\sim 10^{-3}$ m$^{-2}$ yr$^{-1}$, 
        so our conclusion is that the Higgs bosons are important 
        as far as the current indirect dark matter search experiments 
        are concerned.
        The contribution from final states containing Higgs bosons 
        results in a lower bound for the muon flux from the Sun of 
        $10^{-7} \div 10^{-8}$ m$^{-2}$ yr$^{-1}$ 
        for neutralinos heavier than about 200 GeV.
        In other words, if the collider experiments show that the 
        neutralino is so heavy, 
        one should see signals at the latest as the 
        indirect search experiments reach a sensitivity of about 
        $10^{-7}$ m$^{-2}$ yr$^{-1}$.
        And this is due to the contribution from the Higgs final states.

\section{Energetic Tau Neutrinos from Charged Higgs Boson}
        In the last section we saw that the charged Higgs boson 
        can decay directly into $\tau$ neutrinos via 
        $H^+ \rightarrow \tau^+ \nu_{\tau}$, 
        which is the dominant channel below the $tb$ threshold 
        in most cases except for small $\tan \beta$ values.

        The $\tau$ neutrinos as the direct decay product of the charged Higgs  
        from neutralino annihilation can be expected to have 
        the highest energies among all $\tau$ neutrinos from WIMPs 
        and have a harder energy spectrum.
        The charged Higgs boson produced in 
        $\chi \chi \rightarrow W^{\pm} H^{\mp}$ 
        decays with energy between $m_{\chi}$ and
        $2 m_{\chi}$ depending on the charged Higgs mass
        (currently we know that $m_{\rm Ch} \geq 78.6$ GeV
\cite{PDG2000}).
        The maximum energy the $\tau$
        neutrinos can obtain from the charged Higgs boson produced in 
        neutralino annihilation is always about the neutralino mass 
        $m_{\chi}$, independent of the charged Higgs mass $m_{\rm Ch}$.
        The channel $\chi \chi \rightarrow W^{\pm} H^{\mp}$ is open when 
        the kinematic condition $2 m_{\chi} > m_W + m_{\rm Ch}$ is satisfied.
        One thing interesting is that when the charged Higgs mass 
        $m_{\rm Ch}$ not deviates too much from $2 m_{\chi} - m_W$, 
        the $\tau$ neutrino energy spectrum should 
        have the form of a relatively sharp peak
        at half the energy of the decaying charged Higgs
\footnote{
        Indeed, if one assumes $M_W\ll m_{\rm Ch}$
        (or equivalently $M_W\approx 0$),
        after the annihilation $\chi \chi \rightarrow W^{\pm} H^{\mp}$
        both $W$ and Higgs bosons can appear at rest
        and $E_{H^+} =  m_{\rm Ch} \approx 2 m_\chi$. 
        Therefore the decay $H^+ \rightarrow \tau^+ \nu_{\tau}$ gives
        $E_\nu = E_\tau = m_{\rm Ch}/2 = m_\chi$.
        Under the same assumption the competitive annihilation
        $\chi \chi \rightarrow W^{\pm} W^{\mp}$ 
        with  $W \rightarrow \tau \nu$ gives
        $E_{W^\pm} = m_{\chi}$ followed by
        $E_\nu = E_\tau = m_{\chi}/2$.
        Therefore the hardest part of the $\nu_\tau$ energy spectrum one can 
        connect with the decay of the (very) heavy charged Higgs boson.
}
. 
        In contrast, other neutralino annihilation products like 
        fermion and gauge boson pairs decay always with energy equal 
        to $m_{\chi}$ and generate a much 
        flatter $\tau$ neutrino spectrum.
 
        The flux of the $\tau$ neutrinos from the charged Higgs
        bosons produced in WIMPs annihilation and
        decaying with energy and velocity 
        $E_{\rm Ch}$ and $\beta_{\rm Ch}$, respectively, can be calculated 
        using the formulae for gauge boson decays given in
\cite{Jun95} 
\begin{eqnarray}
\left( \frac{d \phi}{dE} \right)^{\oplus}_{H^+} ( E_{\nu} ) 
   &=& \frac{\Gamma_A}{4 \pi R^2} Br (\chi \chi \rightarrow H^{\pm} W^{\mp}) 
   \left( \frac{dN}{dE} \right)_{H^+} ( E_{\nu} )
   \nonumber \\
&&
   \nonumber \\
   &=& 
   \frac{\Gamma_A}{4 \pi R^2} 
   Br (\chi \chi \rightarrow H^{\pm} W^{\mp}) 
   \frac{\Gamma (H^+ \rightarrow \tau^+ \nu_{\tau})}
   {E_{\rm Ch} \beta_{\rm Ch}} \times 
   \\
&& 
   \nonumber \\
&&
   \hspace{4.4cm} \Theta (E^{\oplus}_{\rm low} < E_{\nu} <
   E^{\oplus}_{\rm upp}), \nonumber 
\label{tau-flux-earth}
\end{eqnarray}
        where 
\begin{equation}
   E^{\oplus}_{\rm low} = \frac{E_{\rm Ch}}{2} 
                        (1 - \beta_{\rm Ch}), \hspace{0.4cm}
   E^{\oplus}_{\rm upp} = \frac{E_{\rm Ch}}{2} (1 + \beta_{\rm Ch})
\end{equation}
        are the lower and upper boundaries for the $\tau$ neutrino energy.
        $\Theta (x)$ is the Heaviside step function, 
        $\Theta (x) = 1$ if $x$ is true and 0 otherwise.

\begin{figure}[h]
\begin{center}
\includegraphics[height=11cm]{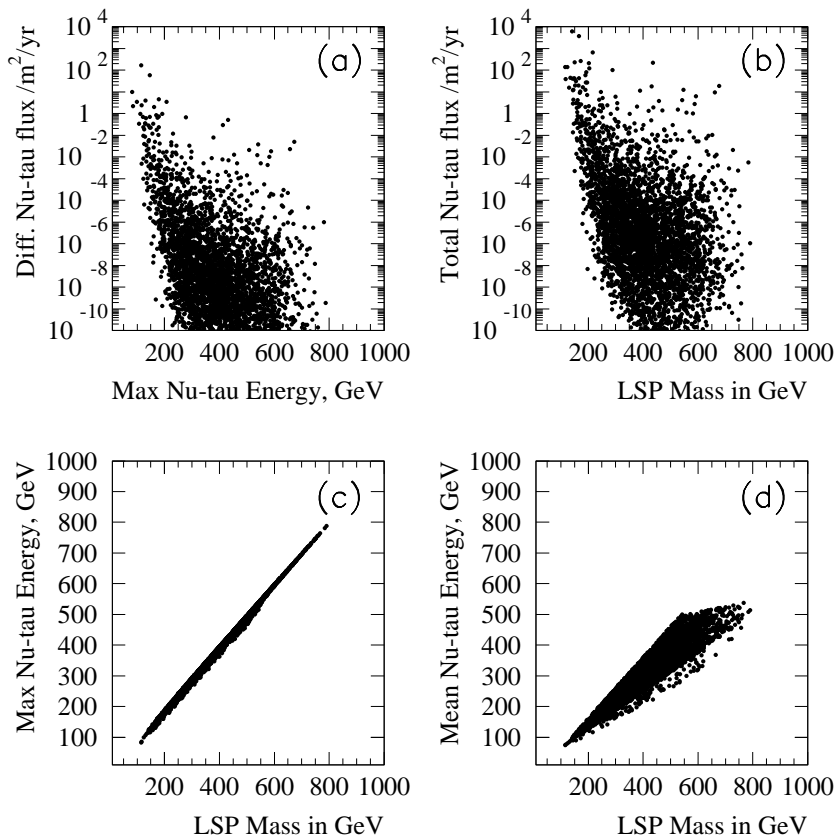}
\end{center}
\caption{(a) Differential flux of $\tau$ neutrinos from charged Higgs 
boson decay vs. maximum $\tau$ neutrino energy.
(b) Total flux of $\tau$ neutrinos from charged Higgs bosons decay vs.
neutralino mass $m_{\chi}$. 
(c) Maximum $\tau$ neutrino energy vs. neutralino mass $m_{\chi}$. 
(d) Mean $\tau$ neutrino energy vs. neutralino mass $m_{\chi}$. 
All for the case of the Earth.}
\label{nutauh}
\end{figure}

        In figure 
\ref{nutauh} we plot the differential $\tau$ neutrino flux from 
        charged Higgs boson decay in the Earth versus the maximum $\tau$ 
        neutrino energy 
        $E^{\oplus}_{\rm upp}$; 
        we also integrate the differential neutrino flux
(\ref{tau-flux-earth})
        over $E_{\nu}$ to obtain the total flux of the
        $\tau$ neutrinos from charged Higgs decay in the Earth and plot it 
        versus the neutralino mass. 
        The dependences of the maximum and mean energy of these $\tau$ 
        neutrinos on the neutralino mass are also shown.  
        The maximum energy a $\tau$ neutrino can obtain 
        from the neutralino annihilation via charged Higgs decay 
        $E^{\oplus}_{\rm upp}$ 
        is roughly the neutralino mass, as expected.

        For the $\tau$ neutrinos produced in the core of the Sun,
        the neutrino interactions with the solar medium play an important role
        in determining the $\tau$ neutrino spectrum at the Sun's surface. 
        We make a rough estimate for this effect following 
\cite{Rit88}
\begin{eqnarray}
   \left( \frac{d \phi}{dE} \right)^{\odot}_{H^+} ( E_{\nu} )
   &=& \frac{\Gamma_A}{4 \pi R^2} \hspace*{1mm} 
   Br (\chi \chi \rightarrow H^{\pm} W^{\mp}) 
   \frac{\Gamma ( H^+ \rightarrow \tau^+ \nu_{\tau} )}
   {E_{\rm Ch} \beta_{\rm Ch}}
   \times \\
&&
   \nonumber \\
&& \hspace*{1.9cm}
   \left( 1 - E_{\nu} \tau_{\nu} \right)^{\alpha_{\nu} - 2}
   \hspace*{1mm}
   \Theta (E^{\odot}_{\rm low} < E_{\nu} < E^{\odot}_{\rm upp}) \nonumber
\end{eqnarray}
with
\begin{equation}
E^{\odot}_{\rm low} \equiv
   \frac{E^\oplus_{\rm low}}{ 1 + E^\oplus_{\rm low} \tau_{\nu}}, 
   \hspace*{0.4cm}
E^{\odot}_{\rm upp} \equiv 
   \frac{E^\oplus_{\rm upp}}{ 1 + E^\oplus_{\rm upp} 
   \tau_{\nu}} 
\end{equation}
        the lower and upper boundaries for the energy of the $\tau$ neutrinos
        from the Sun.
        The coefficients 
        $\tau_{\nu_{\tau}}$,
        $\tau_{\bar{\nu}_{\tau}}$,
        $\alpha_{\nu_{\tau}}$ and
        $\alpha_{\bar{\nu}_{\tau}}$ are used to parametrize the 
        neutrino stopping and absorption effects in the Sun
\cite{Rit88,Mori93}.
        Due to the interactions with the solar medium, the $\tau$ neutrino 
        differential flux from the Sun depends on the 
        neutrino energy $E_{\nu}$ explicitly. 
        We choose to plot $(d \phi / d E)^{\odot}$ versus 
        $E^{\odot}_{\rm upp}$, since the flux 
        is largest for the highest energy.
       
\begin{figure}[t]
\begin{center}
\includegraphics[height=11cm]{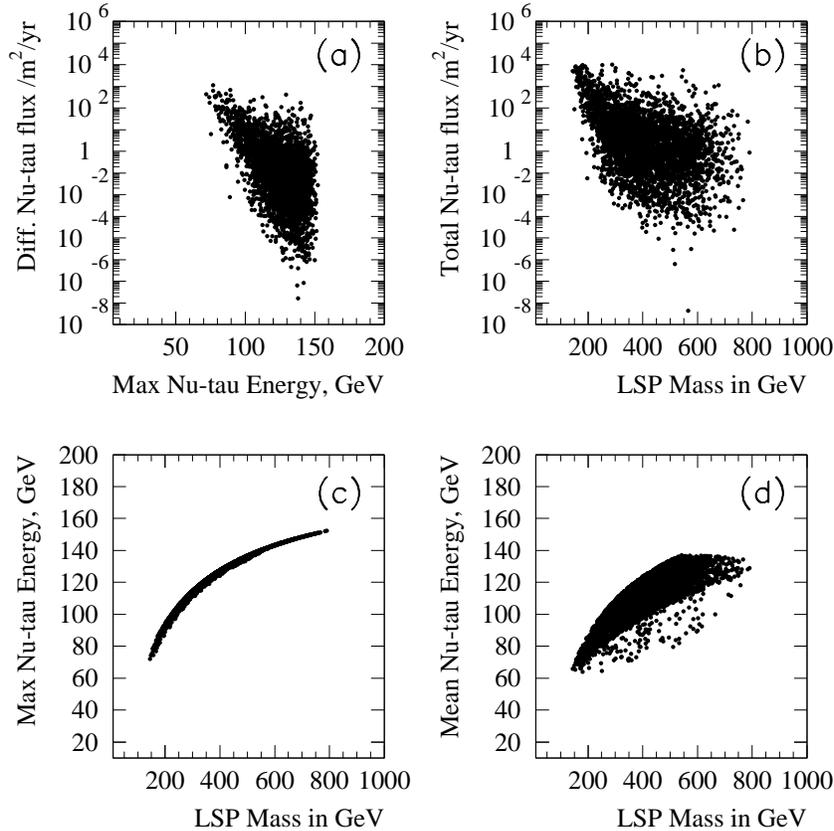}
\end{center}
\caption{(a) Differential flux of $\tau$ neutrinos from charged Higgs 
boson decay vs. maximum $\tau$ neutrino energy.
(b) Total flux of $\tau$ neutrinos from charged Higgs bosons decay vs.
neutralino mass $m_{\chi}$. 
(c) Maximum $\tau$ neutrino energy vs. neutralino mass $m_{\chi}$. 
(d) Mean $\tau$ neutrino energy vs. neutralino mass $m_{\chi}$. 
For the case of the Sun.}
\label{nutauhs}
\end{figure}

        In figure
\ref{nutauhs} we show our results for the $\tau$ neutrinos from 
        the charged Higgs decay in the Sun.
        By comparing it with figure 
\ref{nutauh} we see that the energy of the $\tau$ neutrinos 
        from the Sun is to a large extent degraded due to the 
        interactions with the solar medium, however, since all neutrinos 
        produced by the neutralino annihilation in the core of the Sun 
        suffer from these effects while propagating through the Sun,
        the $\tau$ neutrinos from charged Higgs decay are still the most
        energetic among all neutrinos from neutralino annihilation in the Sun.
        Furthermore, we can expect a much larger  
        $\tau$ neutrino flux from the Sun than from the Earth. 

        All other neutralino annihilation products can also produce $\tau$
        neutrinos.
        But for our purpose only those channels which can produce $\tau$ 
        neutrinos with comparable fluxes in the same energy range as the 
        charged Higgs boson does should be taken into account.
        Therefore, the light fermion pairs need not be considered, 
        since the annihilation of neutralino into them are negligible 
        when the channel $\chi \chi \rightarrow H^{\pm} W^{\mp}$ is open.
        The annihilation into top quark pair is one of the dominant 
        channels for heavier neutralinos.
        Top quark decays almost exclusively via $t \rightarrow W^+ b$ 
        with energy $m_{\chi}$.
        We make an estimate using the formulae given in
\cite{Jun95}
        and found that though the maximum $\tau$ neutrino energy from top 
        quark  decay is also roughly $m_{\chi}$, the differential flux 
        at the maximum energy value is negligible. 
        Furthermore, the energy spectrum is much softer.
        Therefore we do not take into account the contribution from 
        top quark decay when considering energetic $\tau$ neutrinos 
        from charged Higgs decay. 
        Following similar arguments, the neutral Higgs bosons 
        $h^0$, $H^0$ and $A^0$ are also excluded.
        The gauge bosons $W^{\pm}$ and $Z$ decay with energy $m_{\chi}$
        into $\tau$ neutrinos 
        with branching ratios 0.105 and 0.067, respectively.   
        For heavier neutralinos, the maximum energy of the 
        $\tau$ neutrinos from gauge boson decay is also roughly 
        $m_{\chi}$, the mean energy is always $m_{\chi} / 2$. 
        Our calculations show that the $\tau$ neutrino flux 
        from gauge boson decays is comparable to or even greater 
        than that from charged Higgs decay in the same energy range.
        One can expect an enhancement 
        of the total $\tau$ neutrino flux by a factor of 2  
        due to the contributions from gauge boson pairs
        from neutralino annihilation.

        Therefore the $\nu_\tau$ signal from the charged
        Higgs bosons one can in principle find only   
        searching for an enhancement of the 
        $\nu_\tau$ spectrum at highest $\tau$ neutrino energies.        
        The larger the LSP and the charged Higgs boson masses 
        the stronger can be the expected enhancement.

\section{Conclusions}
        In this work we studied the role of the Higgs bosons in the 
        indirect detection of the WIMPs via their annihilation into 
        energetic neutrinos.
        We performed our calculations on the basis of the MSSM 
        parameter space at the electroweak scale, 
        we did not assume any universality relations 
        for the parameters at the unification scale.

        First we have investigated the contribution to the 
        total upward-going 
        muon detection rate from neutrinos produced in the neutralino 
        annihilation final states containing Higgs bosons.
        We found that their contribution is in general significant if the 
        neutralino is heavier than about 150 GeV, 
        and is becoming dominant in the case of the Sun as the current 
        indirect WIMPs detection experiments reach the sensitivities of
        $\sim 10^{-3} \div 10^{-4}$ m$^{-2}$ yr$^{-1}$.
        But most important, the final states containing Higgs bosons 
        result in a lower bound on the expected detection rate 
        from the Sun of 
        $\sim 10^{-7} \div 10^{-8}$ m$^{-2}$ yr$^{-1}$ 
        for neutralinos heavier than about 200 GeV.

        While the neutral Higgs bosons do not decay into muon neutrinos 
        directly,
        the charged Higgs boson produced in the neutralino 
        annihilation channel
        $\chi \chi \rightarrow W^{\pm} H^{\mp}$
        has a large probability to decay into $\tau$ neutrinos 
        directly via
        $H^{\pm} \rightarrow \tau \nu_{\tau}$.
        We expect them to be amongst the most energetic neutrinos 
        from WIMPs annihilations and to have a very hard energy spectrum.
        We have estimated the energy and flux of these $\tau$ neutrinos 
        from the Earth and the Sun, 
        whereby in the case of the Sun the effects of neutrino
        stopping and absorption due to neutrino interactions 
        with the solar medium were taken into account following 
\cite{Rit88}.
        Our results show that the maximum energy of the $\tau$ 
        neutrinos from charged Higgs decay in the Earth is roughly 
        the neutralino mass 
        $m_{\chi}$, and their mean energy is just the energy of the 
        charged Higgs boson, which  depends only on the masses of the 
        neutralino and the charged Higgs.
        The flux is in most cases below 1 m$^{-2}$ yr $^{-1}$ and
        lower for heavier neutralinos. 
        Only for neutralino masses around 
        150--250 GeV one can expect higher $\tau$ neutrino flux.

        In the case of the Sun, the energies of the $\tau$ neutrinos 
        are strongly degraded due the interactions with the solar medium
        (not greater than 150 GeV even for the largest neutralino mass),
        the energy spectrum is also softer. 
        But since all neutrinos propagating from the core of the Sun 
        to the surface suffer from the same effects, 
        the $\tau$ neutrinos from charged Higgs  
        decay in the Sun are still among the most energetic 
        and in principle can be separated when detected.
        Furthermore, one can expect much higher $\tau$ neutrino flux 
        from the Sun, about $10^2 \div 10^4$ m$^{-2}$ yr$^{-1}$, 
        than from the Earth.

        The possible observation of such energetic $\tau$ 
        neutrinos would be an evidence of the
        neutralino annihilation into charged Higgs bosons.  
        In SUSY scenarios with very heavy sfermion and gaugino masses 
        the existence of the charged Higgs boson
        could be considered as a clear signal of the supersymmetry.

\section*{Acknowledgments}
        We thank 
        Prof. Pran Nath and 
        Dr. O. V. Suvorova for the helpful discussions.

\end{document}